\documentclass[twocolumn,aps, prl]{revtex4}

\usepackage{graphics}
\usepackage{graphicx}
\usepackage{dcolumn}
\usepackage{bm}
\usepackage{amsmath,amssymb}

\newcommand{\de}{\partial}

\newcommand{\eq}[2]{\begin{equation} \label{#1} #2 \end{equation}}

\newcommand{\sgn}{\textrm{sgn}}

\newcommand{\etal}{{\em et al.}}

\begin{document}

\title{Diffractive resonant radiation emitted by spatial solitons in waveguide arrays: towards a spatiotemporal supercontinuum generation}
\author{Truong X. Tran$^{1}$ and Fabio Biancalana$^{1,2}$}
\affiliation{$^{1}$Max Planck Institute for the Science of Light, G\"{u}nther-Scharowsky str. 1, 91058 Erlangen, Germany \\
$^{2}$School of Engineering and Physical Sciences, Heriot-Watt University, EH14 4AS Edinburgh, UK}
\date{\today}

\begin{abstract}
We study analytically and numerically a new kind of diffractive resonant radiation emitted by spatial solitons, which is generated in waveguide arrays with Kerr nonlinearity. The phase matching condition between soliton and radiation is derived and agrees well with direct pulse propagation simulations. The folded dispersion due to the Brillouin zone leads to a peculiar anomalous soliton recoil that we describe in detail. A linear potential applied across the array generates the analogue of the Raman self-frequency shift in optical fibers, only now applied to the wavenumber. We demonstrate that it is possible to mimic closely temporal fiber-optical dynamics, unveiling the new effects of wavenumber-supercontinuum generation and the compensation of the 'soliton self-wavenumber shift' by the emitted diffractive radiation. This work paves the way for designing unique optical devices that generate spectrally broad supercontinua with a controllable directionality.
\end{abstract}
\maketitle

\paragraph{Introduction ---} Waveguide arrays (WAs) consisting of identical, equally spaced waveguides, present a unique discrete platform to explore many interesting fundamental phenomena such as discrete diffraction \cite{jones}, discrete solitons \cite{christodoulides,kivshar}, and photonic Bloch oscillations \cite{peschel,pertsch2,lenz,lederer}. In applications, 2D networks of nonlinear waveguides with discrete solitons may be useful for designing signal-processing circuits \cite{christodoulides2}. Recently, WAs have also been used to mimic relativistic phenomena typical of quantum field theory, such as {\em Zitterbewegung} \cite{zitterbewegung}, Klein paradox \cite{klein}, fermion pair production \cite{fermionpairproduction} and the Dirac equation \cite{diracequation}.

The concept of dispersive resonant radiation (DisRR), which emerges due to higher-order dispersion (HOD) terms, has been well studied in the last decade in the temporal case for optical fibers \cite{kuehl,wai,karpman,akhmediev,biancalana}. When an ultrashort pulse is launched into optical fibers, a DisRR due to the phase matching between the fiber and the soliton group velocity dispersion (GVD) generates one or more new frequencies \cite{akhmediev,biancalana}. This DisRR, together with other nonlinear effects such as self- and cross-phase modulation, soliton fission \cite{fission}, and stimulated Raman scattering \cite{srs}, are the main ingredients of the so-called supercontinuum generation (SCG) \cite{dudley,agrawal1}, particularly in highly nonlinear photonic crystal fibers \cite{russell}. SCG is among the most important phenomena in nonlinear fiber optics which has led to a number of important technological advances in various fields, such as spectroscopy and medical imaging \cite{alfano}, metrology \cite{holzwarth}, and the realization of broadband sources \cite{agrawal2}.

\begin{figure}[htb]
  \centering \includegraphics[width=0.45\textwidth]{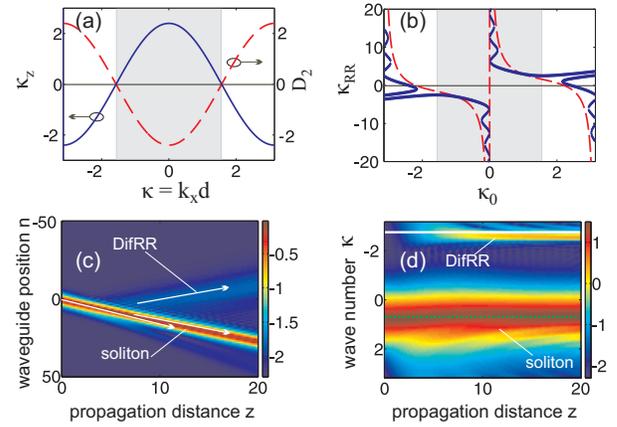}
\caption{\small{(Color online) (a) Solid blue line: WA dispersion $\kappa_{z}$ vs. $\kappa$. Dashed red line: $D_{2}$ vs. $\kappa$, showing the two zero-diffraction points located at $\kappa=\pm\pi/2$. (b) Wavenumber $\kappa_{\rm RR}$ of the generated DifRR, as a function of the input soliton wavenumber $\kappa_{0}$. In both (a,b) the gray shaded area indicates the region where bright solitons can propagate. (c,d) Beam propagation in the ($n,z$)-plane (c) and ($\kappa,z$)-plane (d). Parameters are $A_{0}=0.8$, $C=1.2$, $\kappa_{0}=0.7$, $N=100$.}}
  \label{fig1}
\end{figure}

Inspired by advances in DisRR studies in the last decade, in this Letter we prove theoretically the existence of a new kind of resonant radiation - namely the {\em diffractive resonant radiation} (DifRR), which occurs when a continuous-wave (CW) beam or a relatively long pulse is launched into WAs. Similarities and differences between DisRRs and DifRRs are analyzed. We show that when the phase matching condition is satisfied, a spatial soliton emits DifRR with a new well-defined direction, i.e. transverse wavenumber. Moreover, due to the periodicity of discrete systems, and thus the existence of a Brillouin zone, many unusual effects which cannot exist in continuous media can now occur, such as the anomalous solitonic recoil and the anomalous compensation of soliton self-wavenumber shift, described in this Letter for the first time. This paves the way for producing controlled spatiotemporal supercontinua in both the frequency and wavenumber domain, i.e. broadband radiation with a controlled directionality.

\paragraph{Phase matching condition for the diffractive resonant radiation ---}
Light propagation in a discrete, periodic array of Kerr nonlinear waveguides can be described, in the CW regime, by the following dimensionless equation \cite{kivshar,agrawal1,lederer}:
\eq{CWCM1}{i\frac{da_{n}(z)}{dz} + c[a_{n+1}(z)+ a_{n-1}(z)] + |a_{n}(z)|^{2}a_{n}(z)=0,}
where $a_{n}$ is the electric field amplitude in the $n$th waveguide, $z$ is the longitudinal spatial coordinate, $c\equiv C/(\gamma P)$, where $C$ is the coupling coefficient (in units of a wavenumber) resulting from the field overlap between neighboring waveguides, $\gamma$ is the nonlinear coefficient of a single waveguide, and $P$ is the input beam peak power. By using the stationary discrete plane wave solution for the $n$-th waveguide $a_{n}(z) = a_{0}\exp[i(nk_{x}d +\kappa_{z}z)]$ one arrives, in the linear case, at the dispersion relation between $\kappa_{z}$ and $k_{x}$ \cite{jones}:
\eq{dispersion}{\kappa_{z}(k_{x})=2c\cos(k_{x}d),} where $d$ is the center-to-center spacing between two adjacent waveguides, and $k_{x}$ is the transverse wavenumber, see solid blue line of Fig. \ref{fig1}(a). It is clear from Eq. (\ref{dispersion}) that $\kappa_{z}$ is periodic in $\kappa \equiv k_{x}d$, which represents the phase difference between adjacent waveguides. Thus, within the coupled mode approximation, it suffices to investigate the first Brillouin zone of the folded dispersion, $-\pi \leq \kappa \leq \pi$.

Since a typical input beam has a finite width covering several waveguides, its Fourier spectrum has a certain bandwidth with a central transverse wavenumber $\kappa_{0}$, which is fixed by the input angle of incidence of the exciting beam. We can then use a Taylor expansion of Eq. (\ref{dispersion}) as follows:
\eq{taylor}{\kappa_{z}(\kappa)=\kappa_{z}(\kappa_{0}) + \sum_{m\geq1}\frac{D_{m}}{m!}\Delta\kappa^{m},} where $\Delta\kappa \equiv \kappa - \kappa_{0}$, and $D_{m} \equiv (d^{m}\kappa_{z}/d\kappa^{m})|_{\kappa_{0}}$ is the $m$th order diffractive Taylor coefficient [thus, $D_{1} = -2c\mathrm{sin}(\kappa_{0})$, $D_{2} = -2c\mathrm{cos}(\kappa_{0})$, etc.]. In Fig. \ref{fig1}(a) we plot a typical curve for $D_{2}(\kappa)$ (dashed red line), showing the existence of two zero-diffraction points located at $\kappa=\pm\pi/2$. In a sense, this shape of $D_{2}$ is analogous to the GVD of photonic crystal fibers in the temporal case \cite{russell}.

Following Refs. \cite{pertsch,lederer}, we now approximate the discrete variable $n$ with a continuous one. This is justified since we shall use pulses and solitons that extend for several waveguides. Defining now $n$ as a continuous variable of the distributed amplitude function $\Psi(n,z) = a_{n,z}\exp(-i\kappa_{0}n)$, we eliminate the zero-th order term $\kappa_{z}(\kappa_{0})$ which is responsible for a general phase evolution through the substitution $\Psi(n,z) \rightarrow \Psi(n,z)\exp[i\kappa_{z}(\kappa_{0})z]$. The first order term, $-iD_{1}\de_{n}$, takes into account the transverse velocity and can also be eliminated by introducing a comoving frame, $n\rightarrow n + D_{1}z$. After dropping these two low-order terms one arrives at the following equation:
\eq{mainsimplified}{\left[ i\de_{z} - \frac{D_{2}}{2}\de_{n}^{2} + \sum_{m\geq3}\frac{D_{m}}{m!}(-i\de_{n})^{m} +  |\Psi(n,z)|^{2}\right]\Psi(n,z) =0.}

Equation (\ref{mainsimplified}) is formally identical to the well-known generalized nonlinear Schr\"odinger equation (GNLSE), which describes the evolution of pulses in a single optical fiber, plus HOD terms \cite{agrawal2}. In Eq. (\ref{mainsimplified}) we have the transverse spatial variable $n$ instead of the temporal variable $t$ of the conventional GNLSE. Unlike the temporal GNLSE, where a Taylor series for the fiber dispersion can usually be expanded up to a small number of terms (because HOD coefficients become rapidly very small), in Eqs. (\ref{taylor}) and (\ref{mainsimplified}) many higher-order diffraction terms $D_{m\geq 2}$ should be taken into account, since their absolute values will be either $|2c\mathrm{sin}(\kappa_{0})|$, or $|2c\mathrm{cos}(\kappa_{0})|$, and the sum only converges due to the factorial in the denominator.

In the temporal version of the GNLSE, it is well-known that a temporal soliton propagating in a fiber emits small-amplitude, dispersive and quasi-monochromatic waves at well-defined frequencies (the DisRR) when the linear fiber dispersion and the nonlinear soliton dispersion are matched \cite{akhmediev,biancalana}. It is thus natural to conjecture that, in a WA, a {\em spatial} soliton, which in the continuous variables approximation extends over several waveguides, emits during the propagation a similar kind of small-amplitude diffractive radiation, within a narrow wavenumber range, due to the phase-matching between the spatial soliton nonlinear dispersion and the linear array dispersion given by Eq. (\ref{dispersion}). By using the perturbation approach which was developed for DisRRs in \cite{akhmediev}, here we derive the phase-matching condition for the DifRR in a similar way. We first find the unperturbed soliton solution of Eq. (\ref{mainsimplified}) where all diffractive terms $D_{m\geq3}$ are dropped. The soliton solution of Eq. (\ref{mainsimplified}) is:
\eq{soliton}{a_{\rm sol}(z,n)= A_{0}\mathrm{sech}\left(\frac{nA_{0}}{\sqrt{2c\mathrm{cos}(\kappa_{0})}}  \right)\mathrm{exp}(ik_{\rm sol}z),} where $k_{\rm sol} = A_{0}^{2}/2$ is the spatial soliton longitudinal wavenumber, identical to its temporal counterpart. The bright soliton solution (\ref{soliton}) only exists when $2c\mathrm{cos}(\kappa_{0}) > 0$, i.e. only in half of the Brillouin zone, where $-\pi/2 < \kappa_{0}<\pi/2$. Now we look for the linearized dispersion relation of plane wave solutions of Eq. (\ref{mainsimplified}), by substituting $\mathrm{exp}[i(k_{\rm lin}z + \Delta \kappa n)]$ into Eq. (\ref{mainsimplified}) and using Eq. (\ref{taylor}). We obtain:
\eq{klin}{k_{\rm lin}(\Delta \kappa)\equiv\sum_{m\geq2}\frac{D_{m}}{m!}\Delta\kappa^{m} =2c[\mathrm{cos}(\kappa) - \mathrm{cos}(\kappa_{0}) + \mathrm{sin}(\kappa_{0})\Delta\kappa].} In Eq. (\ref{klin}), $\kappa_{0}$ is the central wavenumber (which is related to the incident angle) of the incident beam, while $\Delta\kappa$ is the detuning from $\kappa_{0}$, and $\kappa = \kappa_{0}+\Delta\kappa$. Energy exchange between radiation and soliton is possible for those values of $\Delta\kappa$ that satisfy
\eq{phasematching}{k_{\rm lin}(\Delta\kappa)=k_{\rm sol},} where $k_{\rm sol}$ is constant and has been defined above. This phase matching condition, an implicit equation for the radiation wavenumber detuning $\Delta\kappa$, is the central result of this Letter, and accurately predicts the DifRR wavenumber as we shall show soon. If we follow what is commonly done for optical fibers, and we take into account only $D_{2}$ and $D_{3}$ in Eq. (\ref{klin}), and ignoring the power dependence ($k_{\rm sol}\rightarrow 0$), one can easily get the approximate DifRR wavenumber in the form $\kappa_{\rm RR}\simeq\kappa'_{\rm RR} \equiv \kappa_{0} + 3/\mathrm{tan}(\kappa_{0})$. Such approximations are perfectly fine in fiber optics when dealing with DisRR -- they lead to very accurate predictions of the DisRR frequency. However, the same approximation is not good enough for the case of WAs, since, as explained above, the coefficients $(D_{m}/m!)$ decay not as rapidly as in the temporal case, and a large number of orders must be taken into account. However, even if not explicit as in the case of the temporal DisRR, Eq. (\ref{phasematching}) is exact and can be easily solved numerically.

\paragraph{Emission of DifRR and soliton anomalous recoil ---} We now prove numerically the formation of DifRR and the accuracy of the predictions made by the phase-matching condition Eq. (\ref{phasematching}).
In Fig. \ref{fig1}(b) we show the DifRR wavenumber $\kappa_{\rm RR}\equiv\kappa_{0}+\Delta\kappa$ as a function of the input soliton wavenumber (which is related to the angle of incidence) $\kappa_{0}$. The blue solid curve is obtained by finding numerically the roots $\Delta\kappa$ of Eq. (\ref{phasematching}), while the dashed red curve shows the approximated analytical expression given in the previous section. It is clear that $\kappa_{\rm RR}'$ is not accurate enough to be used in practice, when compared to the solid line. In Fig. \ref{fig1}(b) we depict the full range of the first Brillouin zone for completeness, but only the interval $-\pi/2<\kappa_{0}<\pi/2$ (indicated by a gray shaded area), in which pulses experience anomalous diffraction, should be considered. Parameters used in Fig. \ref{fig1} are: $A_{0}$=0.8; $c$=1.2; $\kappa_{0}$=0.7. For these parameters, in the range  $0.235< |\kappa_{0}|<\pi/2$, one can find only one solution for $\kappa_{\rm RR}$, but when $0 < |\kappa_{0}|\leq0.235$,  Eq. (\ref{phasematching}) shows several roots  [see the solid blue curve in Fig. \ref{fig1}(b)]. Thus, one should expect to simultaneously generate several DifRRs with different wavenumbers in the latter interval. However, full numerical simulations of Eq. (\ref{CWCM1}) show that only the solution corresponding to the branch that is the closest to the central horizontal axis (i.e., the axis $\kappa_{\rm RR} = 0$) can be generated and observed, and all other DifRR waves corresponding to roots from other branches are too weak to be seen numerically, since the overlap between the soliton spectral tail and the radiation wavenumbers becomes exponentially small. When $\kappa_{0}=0$, i.e. for a direct incidence of the input CW beam, there is no solution for Eq. (\ref{phasematching}), regardless of the parameters used. This is also confirmed by the direct simulation of Eq. (\ref{CWCM1}).

The evolution of a CW beam along $z$ according to Eq. (\ref{CWCM1}) is shown in Fig. \ref{fig1}(c), for an input beam $a_{\rm in}(n)=A_{0}\mathrm{sech}[nA_{0}/\sqrt{2c\mathrm{cos}(\kappa_{0})}]e^{i\kappa_{0}n}$ (i.e. the approximate soliton solution in the continuous limit), and for a WA made of $N=100$ waveguides. After some propagation, around $z\simeq3$, a DifRR is emitted by the soliton. The evolution of the Fourier transform of the field $a(n)$ of Fig. \ref{fig1}(c) along $z$ is shown in Fig. \ref{fig1}(d). The dashed green horizontal line represents the input wavenumber ($\kappa_{0}=0.7$), while the solid white line is obtained by solving Eq. (\ref{phasematching}) numerically, showing excellent agreement with the pulse propagation. In Fig. \ref{fig1}(b) one can notice that the soliton emits the DifRR with a positive detuning $\Delta\kappa$ when $0 < \kappa_{0} < \pi/2$. For instance, when $\kappa_{0}=0.7$, then from Eq. (\ref{phasematching}) one gets $\kappa_{RR}\equiv \kappa_{0}+\Delta\kappa \simeq 3.53$. However, since the Brillouin zone has a limited extension, when $2\pi>\kappa_{RR}>\pi$ the DifRR will be emitted with a negative detuning due to the folding of the band structure. In the example shown in Fig.\ref{fig1}(d) the effective DifRR wavenumber will be equal to $\kappa_{RR} - 2\pi \simeq -2.75$ [see the white solid line in Fig.\ref{fig1}(d)]. This means that in real space the soliton, instead of recoiling in an opposite direction than the DifRR, will recoil {\em towards} the DifRR itself, see the white arrows in Fig. \ref{fig1}(c). The same phenomenon occurs in the wavenumber space: the soliton spectral momentum, instead of recoiling away from the radiation, moves slightly towards it [see Fig. \ref{fig1}(d)]. We call this unique effect {\em anomalous recoil}. The final spectrum of Fig. \ref{fig1}(d) can thus be best described as {\em supercontinuum generation in wavenumber space}.

\begin{figure}[htb]
  \centering \includegraphics[width=0.45\textwidth]{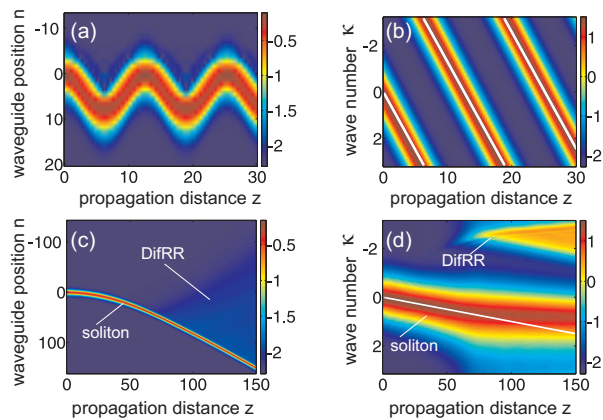}
\caption{\small{(Color online) (a,b) PBOs in the linear regime, shown in both the ($n,z$)-plane (a) and the ($\kappa,z$)-plane (b). Parameters are $c=1$, $\kappa_{0}=0$, $A_{0}=0.8$, $\alpha=0.5$ and $N=35$. (c,d) Anomalous compensation of SSWS in the nonlinear regime depicted in the ($n,z$)-plane (c) and ($\kappa,z$)-plane (d). Parameters are $c=1$, $\kappa_{0}=0$, $A_{0}=0.6$, $\alpha=0.01$ and $N=300$. White solid lines in (b,d) represent the wavenumber calculated by using the formula $\kappa(z) = \kappa_{0} + \alpha z$, making sure that $\kappa\in[-\pi,\pi]$.}}
  \label{fig2}
\end{figure}

\paragraph{Compensation of the soliton self-wavenumber shift ---} The Raman soliton self-frequency shift (SSFS) \cite{mitschke,gordon} is a fundamental effect acting on temporally short pulses in silica fibers, and plays a crucial role in SCG \cite{dudley}. In fibers, SSFS leads to a continuous shift of the central frequency $\omega_{0}$ of the soliton towards the red part of the spectrum \cite{mitschke}. When $\omega_{0}$ is sufficiently close to one of the zero-GVD point of the fiber, DisRR is emitted, and the SSFS can be compensated \cite{skryabinscience,biancalana}.

Here we show that an analogue of the Raman SSFS can be obtained in WAs by applying an external linear potential across the transverse coordinate $n$, which changes the propagation constant along the array in a linear fashion, by using for instance the electro-optic \cite{peschel} or thermal-optic effect \cite{pertsch2}. In the continuous limit, this results in adding a term $\alpha n a_{n}(z)$ to the the left-hand side of Eq. (\ref{CWCM1}), and thus a new term $\alpha n \Psi$ to the left-hand side of Eq. (\ref{mainsimplified}), with $\alpha$ depending on the specific process used to create the potential.
By using the same moment method developed for calculating the shift rate of SSFS in the temporal case \cite{agrawal2,santhanam}, the {\em soliton self-wavenumber shift} (SSWS) rate turns out to satisfy the simple expressions for the central wavenumber and the central position of the soliton,
$\kappa(z) = \kappa_{0} + \alpha z$ and $n_{0}(z)=n_{0}(0)+c \mathrm{cos}(\kappa_{0})(\alpha z^{2} + 2\kappa_{0}z)$, thus the wavenumber shift rate being simply equal to $\alpha$, which is also proportional to the rate of acceleration or deceleration [for $\sgn(\alpha)=\pm1$ respectively] in the $n$ space.

Actually the above formula of $\kappa(z)$ is exact in the absence of Kerr nonlinearity. One immediate effect of the wavenumber shift in the {\em linear} regime is the onset of the well-known photonic Bloch oscillations (PBOs), see Refs. \cite{peschel,pertsch2,lenz,lederer}. We can interpret PBOs in the following way. In Fig. \ref{fig2}(b) a linear beam is launched  in a WA made of $N=35$  waveguides at normal incidence ($\kappa_{0}=0$), but due to the wavenumber shift described above, its wavenumber changes linearly towards positive detunings $\Delta\kappa$ (for $\alpha>0$), and therefore it accelerates towards positive values of $n_{0}$, as shown in Fig. \ref{fig2}(a). However, due to the folding of the band structure, when the wavenumber reaches the value $\kappa=\pi$, it folds back into the first Brillouin zone, and thus $\kappa$ becomes negative, see white solid lines in Fig. \ref{fig2}(b). This means that the beam must change direction, and starts decelerating towards negative values of $n_{0}$, and this process is repeated infinitely in a sinusoidal fashion with a period $z_{0} = 2\pi/\alpha$, see Fig. \ref{fig2}(a). Therefore, the above formula for $n_{0}(z)$ is only valid at the beginning of propagation, before the occurrence of the folding effect.

The analogy between SSFS and SSWS appears instead in the {\em nonlinear} regime. Even though the soliton input wavenumber is far from the zero diffraction points ([located at $\pm\pi/2$ in the dispersion of Fig. \ref{fig1}(a)], eventually the SSWS will push the soliton wavenumber in proximity of one of these two points, depending on the sign of $\alpha$. At this moment, one can conjecture that the soliton will start emitting a strong DifRR that is able to compensate the  SSWS. This is indeed the case, as shown in Figs. \ref{fig2}(b,c). In Fig. \ref{fig2}(c) we show the $z$ evolution of a soliton in the $n$ space with an initial position $n_{0}(0)=0$, in a potential with $\alpha=0.5$. We use a WA made of $N=300$ waveguides, but this is done only in order to make all the figures as clear as possible for the reader, while the concept works for an arbitrary number of waveguides. The soliton starts accelerating (for $\alpha>0$) towards positive values of $n_{0}$ or decelerating (for $\alpha<0$) towards negative values of $n_{0}$ continuously, until at a specific moment (around $z\simeq 70$) it emits the DifRR. In Fig. \ref{fig2}(d) we show the same as Fig. \ref{fig2}(c), but in the wavenumber space. The soliton starts with zero central wavenumber (i.e. normal incidence, $\kappa_{0}=0$), shifts continuously and linearly its wavenumber (the white solid line shows the predicted shift, $\kappa = \kappa_{0} + \alpha z$, for the central wavenumber) until a considerable amount of soliton energy reaches the zero diffraction point $\kappa=\pi/2$, which triggers the emission of a strong DifRR. This emission is so strong that the wavenumber shift of the soliton is compensated, a process that is an exact analogue of the SSFS compensation in fibers \cite{biancalana}.


However, note that in Fig. \ref{fig2}(c,d), an anomalous recoil identical to that of Figs. \ref{fig1}(c,d) occurs: the DifRR should be emitted by the soliton in the positive detunings $\Delta\kappa>0$, but the folding of the Brillouin zone makes the radiation appearing in the negative detunings -- and thus in this case also the soliton central wavenumber shifts towards the DifRR, and not away from it, as it happens in optical fibers.

\paragraph{Conclusions ---} We have discovered the existence and studied the properties of diffractive resonant radiation emitted by discrete spatial solitons in waveguide arrays. Due to the periodicity, several effects such as an anomalous soliton recoil and the anomalous compensation of the soliton self-wavenumber shift, which have no counterpart in continuous systems, are revealed. The analysis of these unusual phenomena could be useful for understanding supercontinuum generation and many other spatiotemporal effects in waveguide arrays, and are applicable to virtually any nonlinear discrete periodic system supporting solitons, therefore making our results very general and of relevance for a number of very diverse communities.

We acknowledge Ulf Peschel for useful discussions. This work is supported by the German Max Planck Society for the Advancement of Science (MPG).

\end{document}